\documentclass[twocolumn,showpacs,preprintnumbers,amsmath,amssymb]{revtex4}
\usepackage{}
\usepackage{graphicx}
\usepackage{dcolumn}
\usepackage{bm}

\begin{document}

\preprint{APS/123-QED}

\title{Quantum coin flipping secure against channel noises}


\author{Sheng Zhang}

\author{Yuexin Zhang}
\email[]{yuexinzhangcmpa@163.com}
\affiliation{Department of Electronic Technology, China Maritime Police Academy, Ningbo 315801, China}

\date{\today}

\begin{abstract}
So far, most of existed single-shot quantum coin flipping(QCF) protocols failed in a noisy quantum channel. Here, we present a nested-structured framework that makes it possible to achieve partially noise-tolerant QCF, due to that there is a trade-off between the security and the justice correctness. It is showed that noise-tolerant single-shot QCF protocols can be produced by filling the presented framework up with existed or even future protocols. We also proved a lower bound of 0.25, with which a cheating Alice or Bob could  bias the outcome.
\end{abstract}

\maketitle


\section{Introduction}
\label{Intro}
Coin flipping is one of the fundamental cryptographic primitives, establishing a random bit between two spatially separated parties, Alice and Bob. The original coin flipping protocol was firstly introduced by Blum in 1981 in the classical setting\cite{Blum81}. Merely three years later, Bennett and Brassard extended the same idea to quantum domain by proposing a, hereafter, well-known quantum protocol, namely "BB84" protocol\cite{BB84}. Not surprisingly, it raised naturally an important theoretical problem that whether there exists a perfect coin flipping protocol in which no party can totally control the outcome? Fortunately, this problem was resolved in more than a decades later by Mayers and Lo and Chau, who proved that perfect coin flipping is impossible even in the quantum setting\cite{Mayers97, LC98}, since it is widely accepted that classical cryptography based on unproven computational complexity assumptions might be cracked by quantum computers.

The first quantum coin flipping(QCF) protocol, i.e., "BB84" protocol, has been found a serious security problem that dishonest Alice can control the protocol entirely, i.e., she is able to achieve a bias of 0.5, by performing a remote steering attack\cite{BB84}, though it is elegant as a template for some subsequent variants\cite{ATVY00, Ambainis04, BBBG09, AMS10}. Hence, Aharonov et al. announced an improved one(ATVY), aiming to vanquish the remote steering attack, based on a quantum bit escrow protocol\cite{ATVY00}. However, the bias of their protocol, of which the bound is proved up to be 0.354\cite{SR02QIC}, is not desirable yet. Later, Ambainis using qutrits (more than one qubit) devised a new QCF protocol whose bias is greatly reduced to 0.25\cite{Ambainis04}. In addition, Spekkens and Ruldolph independently proved that fair QCF protocols with the same bias, 0.25, can be built upon a restricted class of quantum bit commitment protocols\cite{SR01}. Further, Kitaev pushed this bound down to $(\sqrt{2}-1)/2\approx 0.21$ using semidefinite programms(SDP)\cite{Kit03}. A few years later, his result was substantiated by Chailloux and Kerenidis\cite{CK09}, who announced a QCF protocol following from the construction of using a weak QCF protocol described by Mochon\cite{Moc07} as a subroutine, and proved that its bias is arbitrarily close to Kitaev's bound. In 2011, H\"{a}nggi and Wullschleger have proved new tight bounds of cheating probability $p$ in both classical and quantum cases, for instance, $p=\sqrt{(1-H)/2}$ corresponds to the quantum settings, where $H$ denotes the honest abort probability\cite{hanggi2011tight}.

Above work mostly contribute to strong quantum coin flipping, in which the flavor of each party is not restricted by a fixed outcome. We also have a weak version of QCF \cite{SR02PRL, Amb02, KN04, Moc04, Moc05} differing from the strong one only in that the preferences of the parties are known to each other in advance, a more strict definition of these two variants is presented in Refs.\cite{Moc07} and \cite{CK09}. One of the most significant motivations of initiating the study of weak QCF is to achieve perfect coin flipping, which is proved disappointed by its strong versions. Most generally, the bias of a weak QCF protocol is lower than its counterpart, since one can trivially unbalance a strong QCF scheme such that the bias of the desired outcome can be lowered at the expense of the undesired one\cite{SR02PRL}. Remarkably, Spekkens and Ruldolph described a protocol whose bias is at most $(\sqrt{2}-1)/2$. Later, it is tightened by Mochon who declared a protocol achieving a bias of 0.192\cite{Moc04,Moc05}, and thereafter using Kitaev's formalism, namely semidefinite programming, proved that the bias of weak QCF can be made arbitrarily small\cite{Moc07}.

Interesingly, QCF has also been fertilised by new paradigms with focus on the multiparty scenario. Ambainis et al. investigated multiparty QCF by including both weak QCF and strong QCF as the components, hence the protocol seems more likely to be a tournament in which the parties are divided into pairwise subgroups\cite{ABDR04}. It is also further studied in real-life applications, such as the leader election\cite{Ganz09}.

It is known that real-world quantum cryptography systems, e.g., the intriguing quantum key distribution, have always been suffering from some imperfections\cite{Gisin06Trojan,Makarov06}. Indeed, there is no exception for QCF. In other words, QCF is also sensitive to some deviations from its theoretical model, such as the noise. With the motivation of implementing a real-life QCF system, quantum bit-string generation has been proposed to minimize the unexpected effect of noise\cite{BM04a,BM04b}. However, Berl\'in et al. argue that the same goal can be achieved with purely classical means\cite{BBBG09}. It still leaves open that whether there exist nontrivial single-shot QCF protocols of which the expected performance can be achieved in the presence of noise?

Of particular attention, loss is an even bigger threat to a practical QCF implementation. Differing from that in quantum key distribution, it is not difficult for a dishonest party to take advantage of it for a malicious purpose without even being caught cheating. In fact, any QCF protocol is such a cunning two-party game that a loss-tolerant protocol is always obtained at a price. QCF was first reviewed in the loss-tolerant aspect by Berl\'in et al. who constructed their protocol (BBBG09)upon the BB84 template, the price for a loss-tolerant characteristic is that the bias increased dramatically to 0.4\cite{BBBG09}. They also experimentally verified their protocol in practical quantum channels\cite{berlin2011experimental}, as is a great progress upon the first QCF experiment by G.Molina-Terriza et al. in 2005\cite{molina2005experimental}. Subsequently, Aharon et al. presented a class of loss-tolerant protocols without the use of bit commitment, and the bias can be reduced, though not so remarkably\cite{AMS10}. With the same purpose, Chailloux recently announced a new protocol in which a classical encryption step is added to Berl\'in et al.'s, and it approximately produces a bias of 0.359\cite{Chailloux11}. Later, Anna Pappa et al. proposed a practical QCF protocol\cite{pappa2011practical} regarding all possible environmental imperfections including loss, noise and the photon source. Their protocol also uses the template of BBBG09 protocol, except that they replace the photon source with a weak pulse laser. Recently, it is reported that they succeeded in experimentally carrying out a plug-and-play QCF scheme using commercial devices\cite{pappa2014experimental}.

In this paper, we continue to answer the question in Ref.\cite{pappa2011practical} that how to develop new ways to reduce the effect of noise on the honest abort probability. We present a nested structure, with which one can produce a serious of partially noise-tolerant QCF protocols, only with an acceptable price of justice correctness.  The rest of paper is organized as follow: In Sec.\ref{frame}, a framework, referring to the nested structure, is introduced. Then, the security is investigated conditioned with channel noises and losses in Sec.\ref{per}, and a lower bound is also obtained. In the following section, the justice correctness is discussed. At last, a conclusion is drawn.

\section{Framework}
\label{frame}

It is necessary to formalize a quantum noisy channel before we present the framework. A natural way to describe the dynamics of an open quantum system is to regard it as arising from an interaction between the system of interest, which we shall call the principal system, and an environment, which together form a closed quantum system. It is obvious that a quantum channel is inevitable of a loss as well as noise in practice, thus we introduce a parameter $\eta$ to quantify the degree of the loss. Since either party can replace a perfect channel or detection apparatus, we set $\eta=1$.

We define $\rho_{env}=|0\rangle\langle0|$ as the initial state of the environment in a three-level quantum system $\mathcal{H}_{e}$. The environment quantum state transform can be written as

\begin{equation}
\varepsilon(\rho)=Tr_{B}(U_{e}(\rho_{0(1)}\otimes|0\rangle\langle0|)U^{\dag}_{e}).
\end{equation}

Here, $\rho_{x}$ is the quantum state Alice prepares in a Hilbert space $\mathcal{H}_{A}$, and the subscript $x$($x=0\ or\ 1$) stands for the encoding of the state. $U_{e}$, which describes the noisy channel, is a unitary operator acting on the composite system $\mathcal{H}_{A}\otimes \mathcal{H}_{e}$.

Bob measured the received qubit $\rho_{0(1)}$ in a basis randomly chosen from the ones, in which Alice prepares her own qubits. This can be described by a POVM operator $\Pi_{m}$. Therefore, the error rate denoted by $p_{\hat{x}}^{e}$ is written as

\begin{equation}\label{errate}
p_{0(1)}^{e}=Tr[\Pi_{1(0)}\varepsilon(\rho)]
\end{equation}

where $\sum_{0,1}\Pi_{i}=I$.

Since there is a real channel noise and considering a honest Alice, if Bob measures a different result in comparison with Alice's one, he has no reason to abort the protocol. On the contrary, Alice may cheat Bob like this: She simply declares her desired bit to Bob after he has revealed his classic bit. Consequently, if Bob's measurement is not corresponding with Alice's announcement, Alice would blame it to a imperfection of the quantum system. In other words, Bob is unable to verify between a cheating strategy and a truth. Here, to generalize the common property of noise-based cheating, we define the notion that one can not verify the truth of the other party's word as 'Blinding Area'(\emph{BA}).

Following the same idea, Alice might also take advantage of \emph{BA} to perform the noise-based cheating as follow: By the time she has to reveal the classical information, she could announce whatever bit she likes with no risk. \emph{BA} occurs when Bob's measurement result is inconsistent with Alice's information, i.e., Bob cannot tell whether it is resulted from the noise or Alice's intention. Consequently, Bob may abort the protocol in case of judging a honest Alice as a cheater due to a genuine error, as is referred to an event of justice error.

\begin{figure}
  \includegraphics{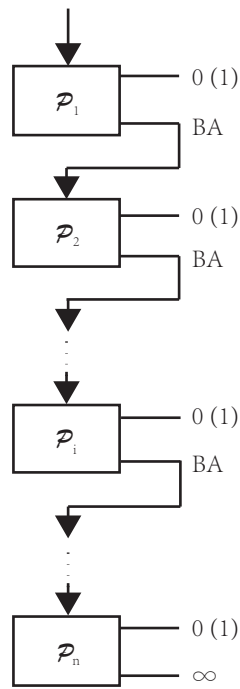}\\
  \caption{Schematics of our framework. It is composed of $n$ elements, denoted as $\mathcal{P}_{i}$ ($i=1,2,...,n$). Here, \emph{BA}, for example in $\mathcal{P}_{i}$, acts an output port which produces a "true" result, which immediately triggers the next element $\mathcal{P}_{i+1}$, only if \emph{BA} takes places there. Otherwise, protocol $\mathcal{P}_{i}$ outputs an outcome coin 0 or 1. $\infty$ at the end of the picture indicates the outcome as "failure".}\label{Fram}
\end{figure}

The central problem that existed single-shot QCF protocols failed in a noisy channel is all due to the presence of \emph{BA}. Therefore, we are motivated to exploit a new technique to reduce its effect. Here, we present a framework initialized by the idea of weighing the noise-tolerance with the justice correctness. As is showed in Fig. \ref{Fram}, a nested structure, in which hundreds of QCF protocols are correlated, corresponds to the presented framework. Each element protocol $\mathcal{P}_{i}$ ($i=1,2,..., n$) is chosen from existed or even future QCF protocols, which might be loss-tolerant as well. Protocol $\mathcal{P}_{i}$, triggered by $\mathcal{P}_{i-1}$ where \emph{BA} occurs, produces an outcome coin(0 or 1) with a certain probability, only if it is in absence of \emph{BA}. Notice that the framework might fail in outputting a legitimate outcome at the $n_{th}$ level if \emph{BA} occurs in $\mathcal{P}_{n}$.
Also, parameter $n$ should be fixed before Alice and Bob begin to run the framework, since it is important to bound the biases of both Alice and Bob.

\section{Performance}
\label{per}

Before we begin to prove a maximum bias for Alice or Bob, it is necessary to summarize some definitions. The QBER denoted by $0\leq P_{e}\leq1$, given by $p_{\hat{x}}^{e}$ from Eq.\ref{errate}, is used to quantify the degree of error. We denote by $Prob\{T_{A},H_{B}\}$ the successful probability for a cheating Alice using a tricky strategy $T_{A}$ while Bob is honest. Therefore, an ideal QCF protocol enjoys an equal success probability for both parties who are honest, i.e.,

\begin{equation}
Prob\{H_{A},H_{B}\}=\frac{1}{2}.
\end{equation}

If one of them uses cheating strategy and the rest remains honest, it can be written as

\begin{equation}
\forall T_{A}\ Prob\{T_{A},H_{B}\}\leq p,
\end{equation}

\begin{equation}
\forall T_{B}\ Prob\{H_{A},T_{B}\}\leq q,
\end{equation}
where $p$ is the maximal probability for cheating Alice and $q$ for Bob. If we implement a QCF protocol in a practical quantum channel, the security criteria is given by the following two inequalities,

\begin{equation}
\forall T_{A}\ Prob\{H_{A},T_{B},P_{e}\}\leq P_{A}.
\end{equation}

\begin{equation}
\forall T_{B}\ Prob\{H_{A},T_{B},P_{e}\}\leq P_{B}
\end{equation}

After introducing channel noises, the cheating probability for Alice immediately turns to be

\begin{equation}\label{8}
P_{A}=p_{*}+(p-p_{*})\times(1-P_{e})+\overline p\times P_{e},
\end{equation}
where $p_{*}$ is the probability that Bob cannot verify Alice's commitment, for example when his measurement basis is not consistent with Alice's.

For Bob, we have
\begin{equation}\label{9}
P_{B}=q\times(1-P_{e})+\overline q\times P_{e}.
\end{equation}
Obviously, Eqs.\ref{8} and \ref{9} imply that Both Alice and Bob performed the same strategies that they ever used in a noiseless scenario.

\subsection{Maximal Biases}

Conventionally, one may presume that a cheating party, Alice or Bob, is powerful enough to do anything only constrained by quantum mechanics. Therefore, the noisy channel may be replaced by the cheater with a noiseless one in advance, as is a natural way to maximize the bias. Here, we shall generalize the biases as functions of the noise rate $P_{e}$, to make a better interpretation of the framework performance, maximal ones are just obtained  by letting $P_{e}=0$. Moreover, only the case of $P_{e}<0.5$ is taken into account. For $P_{e}>0.5$, a clever Alice would always reveal the opposites of her original bits, otherwise she will be a victim of channel noises. In other words, the probability, that a honest Bob will obtain a measurement result consisted with the original a honest Alice reveals, is only $1-P_{e}$, which is lower than 0.5. Therefore, it is reasonable to consider only a case of $P_{e}<0.5$, according to this symmetry.

Interestingly, it is natural from Eqs.\ref{8} and \ref{9} that any QCF protocol involves to a perfect one while $P_{e}=0.5$, i.e.,
\begin{equation}
P_{A}=P_{B}=0.5.
\end{equation}
In this case, any QCF protocol seems to be an "empty" scheme, since the randomness of the outcome coin is totally controlled by the channel.

Using classical probability theory, we have calculated the biases, given by Lemma 1 and Lemma 2. Before this, we should define $\overline{P_{A_{0}}},\overline{P_{B_{0}}}\equiv1$, for Alice and Bob, respectively. Also, we need a further assumption that the optimal cheating strategy, which Alice or Bob performs on the presented framework, is a combination of the ones on each element $\mathcal{P}_{i}$ in a noiseless channel.

\emph{Lemma} 1: Given $n=N$ for the presented framework, Alice's bias $\varepsilon_{A}$ is calculated by

\begin{equation}
P_{A}^{(N)}=\varepsilon_{A}+0.5=\sum^{N}_{i=1}P_{A_{i}}\prod^{i-1}_{j=0}\overline{P_{A_{j}}}.
\end{equation}

\emph{Proof} For $n=1$, we have $P_{A}^{(1)}=P_{A_{1}}$, where $P_{A_{1}}$ is the successful probability that cheating Alice biases the coin to her like in protocol $\mathcal{P}_{1}$. For the case of $n=2$, it is easy to find
\begin{equation}\label{12}
P_{A}^{(2)}=P_{A_{1}}+\overline{P_{A_{1}}}\times P_{A_{2}}.
\end{equation}
$\overline{P_{A_{1}}}$ corresponds to the probability that \emph{BA} took place in $\mathcal{P}_{1}$. Thus, $\mathcal{P}_{2}$, triggered by \emph{BA} in $\mathcal{P}_{1}$, produces an outcome with a probability of $\overline{P_{A_{1}}}\times P_{A_{2}}$. With the same technique, the probability that $\mathcal{P}_{n}$($n=3,4,..., N$) produces an outcome is given by $P_{A_{n}}\prod^{n-1}_{i=1}\overline{P_{A_{i}}}$. Accounting with all cases, one immediately has
\begin{equation}
P_{A}^{(N)}=P_{A_{1}}+P_{A_{2}}\overline{P_{A_{1}}}+\cdots+P_{A_{N}}\prod^{N-1}_{i=1}\overline{P_{A_{i}}},
\end{equation}
which concludes the proof.

Following the same template, we have lemma 2 to calculate Bob's bias.

\emph{Lemma} 2: Given $n=N$ for the presented framework, Bob's bias $\varepsilon_{B}$ is calculated by
\begin{equation}
P_{B}^{(N)}=\varepsilon_{B}+0.5=\sum^{N}_{i=1}P_{B_{i}}\prod^{i-1}_{j=0}\overline{P_{B_{j}}}.
\end{equation}

\emph{Corollary} 1: The framework cannot be totally controlled by any cheating party, given $\mathcal{P}_{i}$ is a secure QCF protocol for all $i$s, i.e.,
\begin{equation}
P_{A}^{(N)}<1, P_{B}^{(N)}<1
\end{equation}

See Appendix for the proof.

From Lemmas 1 and 2, it is seen that $P_{A}^{(N)}$ and $P_{B}^{(N)}$ simultaneously increase as $N$ grows up, since we have $P_{A(B)}^{(n)}-P_{A(B)}^{(n-1)}=P_{A(B)_{n}}\prod^{n-1}_{i=1}\overline{P_{A(B)_{i}}}>0$. In other words, the security of the presented framework degrades as $N$ increases. However, it is possible that the framework ends up with a justice error, that a honest Alice is caught cheating due to the presence of noise taking place in the $N_{th}$ round referred to $\mathcal{P}_{N}$.

\subsection{Justice Error}
Next, we shall investigate the justice error of our framework, in response to the former claim "weighing the noise-tolerance with the justice correctness". Since justice error only occurs in a situation where Alice is honest and Bob is able to verify her commitment,  it is not difficult to compute the justice error rate by
\begin{equation}\label{16}
P_{sys}=\prod_{i=1}^{N}(1-p_{*}^{(i)})P_{e}^{N},
\end{equation}
where $P_{sys}$ denotes the rate.

Obviously, the more frequently we restart the protocol, the smaller $P_{sys}$ is, and it reaches to zero when $N$ approaches infinity. In this case, the framework is nearly broken, while the justice error never occurs. Interestingly, partial noise-tolerance property is achieved with an acceptable loss of justice correctness when $N$ is set finite. Let $p_{*}=0.5$, referring to protocols in which only two nonorthogonal basis are employed, we have $P_{sys}\leq 0.25$ for $N=1$ and $P_{sys}\leq 0.063$ for $N=2$, given $P_{e}$ previously assumed to be less than 0.5. Therefore, it is happy to see that $N=2$ is sufficient enough to reduce the effect of justice errors.

Note that justice error is a systematic price for the noise-tolerant property, yet it repays us the expected security of most single-shot QCF protocols which were supposed to fail in noisy channels.

\subsection{Optimal Situation}

It is clear from above that there is a trade-off between the security and the justice correctness, although the presented framework is partially noise tolerant. Of course, parameter $N$ can be forced to be 1, i.e., Bob firmly calls Alice a cheater in case of $x\neq\hat{x}$ despite it is possibly due to the channel noises. Consequently, the probability of being betrayed by a real noise for a honest Alice is up to $0.25$, which is likely to be unacceptable in real-life applications. In addition, it shouldn't be neglected that in real life implementation, a cheating party is presumed to have unlimited capacity, including replace the noisy channel with a noiseless one to maximize his or her benefit. We have loosely plotted $P_{A}^{(N)}$ as a function of $P_{e}$ with prefixed parameters in Fig2. It is showed that a cheating Alice will gain the most benefit from the channel noises by replacing the channel, i.e., $P_{e}=0$. Similarly, Bob is encouraged to do so before he decides to cheat Alice.

\begin{figure*}
  \includegraphics{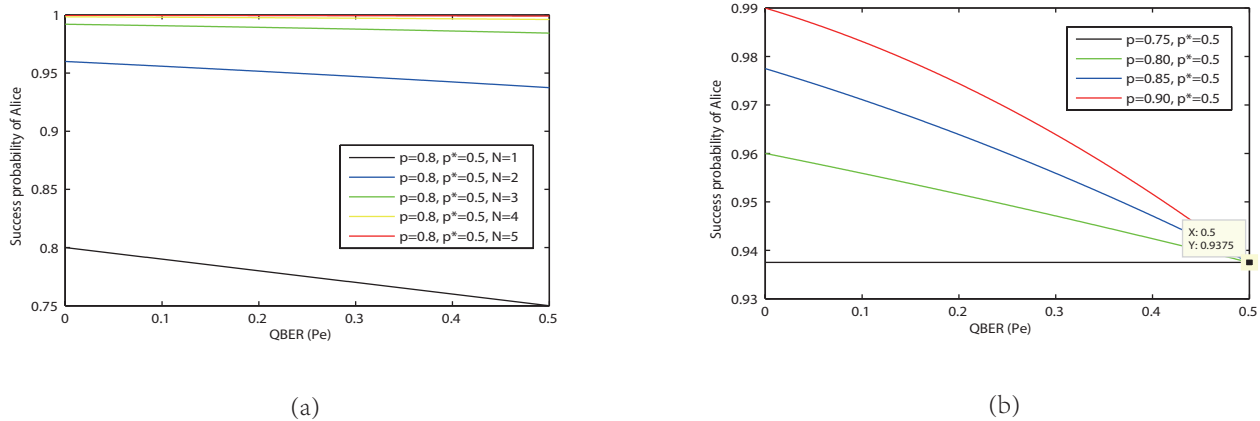}\\
  \caption{Successful probability of Alice as a function of $P_{e}$. (a)The curves are plotted with different $N$s, by setting $p^{\ast}=0.5$ and $p=0.8$ for each element $\mathcal{P}_{i}$, i.e.,each element $\mathcal{P}_{i}$ shares the same cheating probability. (b)The curves are plotted with different $p$s, and for each we set $N=2$. The date point labeled by a rectangle box is directly explained by Eq.\ref{8}. }\label{principle}
\end{figure*}

In an ideal case, where each element protocol $\mathcal{P}_{i}$ is perfect, i.e., $P_{A_{i}}=0.5$ and $P_{B_{i}}=0.5$ adapted from the argument of H\"{a}nggi and Wullschleger, optimal biases of the presented framework is achieved. Table 1 is a list of the results for the ideal case in specific $N$s, it is showed that the security is increasingly discounted with $N$ becoming larger. Even the protocol is only allowed to be restarted once, the increase of successful cheating probabilities for both Alice and Bob reach up to 0.25. It can be inferred that the bias is approximately approaches 0.5 with $N$ being sufficiently large, yielding that our framework is completely broken. Combining with the justice error rate, it is good to set $N=1$ when $P_{e}$ is far less than 0.5. Otherwise, one should consider to restart the protocol when \emph{BA} occurs, i.e., let $N=2$.

\begin{table}
\centering
\caption{Some results referring to the ideal cases}\label{tab1}
\begin{tabular}{cccc}
\hline
\hline
$N$  &  $\mathcal{P}_{i}$ &  $P_{A(B)}^{(N)}$ &  $\varepsilon_{A(B)}$\\
\hline
2    &$0.50$    & $0.7500$    & $0.2500$\\
3    &$0.50$   & $0.8750$    & $0.3750$\\
4     &$0.50$    & $0.9375$   & $0.4375$\\
5    &$0.50$     & $0.9688$   & $0.4688$\\
6     &$0.50$     & $0.9844$   & $0.4844$\\
\hline
\hline
\end{tabular}
\end{table}

If we fill the framework with BBBG09 protocol and a perfect QCF protocol, for example, Lemmas 1 and 2 produce the following two inequalities

\begin{equation}
P^{(2)}_{A}<\frac{3}{4}+\frac{1}{4}\alpha\beta+[1-(\frac{3}{4}+\frac{1}{4}\alpha\beta)]\times 0.5,
\end{equation}

\begin{equation}
P^{(2)}_{B}<\alpha^{2}+(1-\alpha^{2})\times 0.5.
\end{equation}
Consider a fair scenario, we have
\begin{equation}
\frac{3}{4}+\frac{1}{4}\alpha\beta+[1-(\frac{3}{4}+\frac{1}{4}\alpha\beta)]\times 0.5=\alpha^{2}+(1-\alpha^{2})\times 0.5,
\end{equation}

subject to $\alpha^{2}+\beta^{2}=1$. Solving this system yields

\begin{equation}
\alpha^{2}=0.9, \beta^{2}=0.1,
\end{equation}
which immediately conclude
\begin{equation}
\varepsilon_{A}=\varepsilon_{B}=0.45.
\end{equation}
Replacing BBBG09 protocol with Chailloux's scheme\cite{Chailloux11}, we obtain a fair bias of $0.4295$.

\section{Conclusion}\label{conclusion}
We have solved a practical problem in implementing single-shot QCF protocols in real-life channels, we proved that partial noise-tolerance property is achievable using a nested-structured framework, at an acceptable price. However, the bias produced by filling the framework up with existed loss-tolerant QCF protocols is still unsatisfying, it is still meaningful to further develop new loss-tolerant QCF protocols with smaller biases. In addition, is it possible to introduce our framework to weak QCF protocols? Our future work will be focused on above concerns.

\section{Acknowledgements}
This work was supported by National Natural Science Fundation of China with the project number 60872052.

\section{Appendix of corollary}

According to Lemma 1, we have

\begin{equation}
P^{(N)}_{A}\leq \sum^{N}_{i=1}P_{A_{i}}\prod^{i-1}_{j=0}\overline{P_{A_{j}}},
\end{equation}

which can be rewritten as

\begin{equation}
\begin{array}{rl}
P^{(N)}_{A}\leq &\sum^{N-2}_{i=1}P_{A_{i}}\prod^{i-1}_{j=0}\overline{P_{A_{j}}}\\
&+{P_{A_{N-1}}}\prod_{j=0}^{N-2}\overline{P_{A_{j}}}\\
&+{P_{A_{N}}}\prod_{j=0}^{N-1}\overline{P_{A_{j}}}.
\end{array}
\end{equation}

Consequently, it equivalently evolves to

\begin{equation}
\begin{array}{rl}
P^{(N)}_{A}\leq &\sum^{N-2}_{i=2}{P_{A_{i}}}\prod^{i-1}_{j=0}\overline{P_{A_{j}}}\\
&+({P_{A_{N-1}}}+\overline{P_{A_{N-1}}}{P_{A_{N}}})\prod_{j=0}^{N-2}\overline{P_{A_{j}}}.
\end{array}
\end{equation}

 Given that $\frac{1}{2}<{P_{A_{i}}}<1$ is true for all $i$s, hence

\begin{equation}
{P_{A_{N-1}}}+\overline{P_{A_{N-1}}}{P_{A_{N}}}<1.
\end{equation}

Thus, the following inequality also holds
\begin{equation}
\begin{array}{rl}
P^{(N)}_{A}&<\sum^{N-2}_{i=1}{P_{A_{i}}}\prod^{i-1}_{j=0}\overline{{P_{A_{j}}}}+\prod_{j=0}^{N-2}\overline{{P_{A_{j}}}}\\
&=\sum^{N-3}_{i=1}{P_{A_{i}}}\prod^{i-1}_{j=0}\overline{{P_{A_{j}}}}+{P_{A_{N-2}}}\prod^{N-3}_{j=0}\overline{{P_{A_{j}}}}\\
&+\overline{P_{A_{N-2}}}\prod^{N-3}_{j=0}\overline{P_{A_{j}}}\\
&=\sum^{N-3}_{i=1}{P_{A_{i}}}\prod^{i-1}_{j=0}\overline{P_{A_{j}}}+\prod^{N-3}_{j=0}\overline{P_{A_{j}}}.
\end{array}
\end{equation}

Using the same technique, we have
\begin{equation}
\begin{array}{rl}
P^{(N)}_{A}&<\sum^{N-4}_{i=1}{P_{A_{i}}}\prod^{i-1}_{j=0}\overline{{P_{A_{j}}}}+\prod_{j=0}^{N-4}\overline{{P_{A_{j}}}}\\
&...\\
&<\sum^{1}_{i=1}{P_{A_{i}}}\prod^{i-1}_{j=0}\overline{{P_{A_{j}}}}+\prod_{j=0}^{1}\overline{{P_{A_{j}}}}\\
&={P_{A_{1}}}+\overline{{P_{A_{1}}}}\\
&=1.
\end{array}
\end{equation}

Following the same steps, one could also obtain $P^{(N)}_{B}<1$.

\bibliography{my_bib_database}

\end{document}